\documentclass[twocolumn,amsmath,amssymb,prx]{revtex4}
\usepackage{latexsym}
\usepackage{graphicx}% Include figure files
\usepackage{color}
\begin{document}
\title{Superconducting proximity effect in epitaxial Al-InAs heterostructures}

\author{William~Mayer$^{1}$}
\author{Joseph~Yuan$^{1}$}
\author{Kaushini~S.~Wickramasinghe$^{1,2}$}
\author{Tri~Nguyen$^{3}$}
\author{Matthieu~C.~Dartiailh$^{1}$}
\author{Javad~Shabani$^{1}$}
\affiliation{$^{1}$Center for Quantum Phenomena, Department of Physics, New York University, NY 10003, USA
\\
$^{2}$Department of Physics, University of Maryland, College Park, MD 20742, USA
\\
$^{3}$ Department of Physics, City College of City University of New York, New York, NY 10031, USA}

\date{\today}
\begin{abstract}

Semiconductor-based Josephson junctions provide a platform for studying proximity effect due to the possibility of tuning junction properties by gate voltage and large-scale fabrication of complex Josephson circuits. Recently Josephson junctions using InAs weak link with epitaxial aluminum contact have improved the product of normal resistance and critical current, $I_cR_N$,  in addition to fabrication process reliability. Here we study similar devices with epitaxial contact and find large supercurrent and substantial product of $I_cR_N$ in our junctions. However we find a striking difference when we compare these samples with higher mobility samples in terms of product of excess current and normal resistance, $I_{ex}R_N$. The excess current is negligible in lower mobility devices while it is substantial and independent of gate voltage and junction length in high mobility samples. This indicates that even though both sample types have epitaxial contacts only the high-mobility one has a high transparency interface. In the high mobility short junctions, we observe values of $I_cR_N/\Delta \sim 2.2$ and $I_{ex}R_N/\Delta \sim 1.5$ in semiconductor weak links.

\end{abstract}
\maketitle

%Epitaxial two-dimensional heterostructures containing indium arsenide (InAs) layers are hypothesized to be a suitable system for spintronics applications \cite{ZuticRevModPhys}.  Spintronics favors materials like InAs with its strong spin orbit interaction (SOC) and large g-factor.  
Recently, realizing transparent contact in superconductor-semiconductor (S-Sm) systems has become the focus of renewed theoretical and experimental attention partly because of the potential applications in spintronics, topological superconductivity, \cite{Aliceareview,MajoranaReview}, and superconducting quantum computation \cite{Maxim17, Karl17, Larsen15}.  Generally materials considered for S-Sm systems, such as GaAs two-dimensional electron gas(2DEG) \cite{Wan2015} contacted with either aluminum or niobium based superconductors, have had high quality 2DEG's but suffered from imperfect interfaces due to the presence of a Schottky barrier. Narrow bandgap materials such as InAs and InSb have been studied due to the potential for transparent metallic interfaces \cite{Clark1980}, first using 2DEG's \cite{Richter1999,Kroemer1994,Takayanagi95} and more recently using nanowires \cite{Doh2005}. Recently it has been shown that epitaxial contacts to nanowires and near surface quantum wells can improve the proximity effect in Josephson junctions \cite{Chang2015, Shabani2016}. The figure of merit $I_{c}R_{N}$, where $I_{c}$ is the critical current and $R_{N}$ is normal resistance, normalized by $\Delta_0/e$ up to 0.7 has been achieved. These improvements over earlier studies  \cite{Heida99} were associated with the in-situ growth of epitaxial superconducting contact. These epitaxial contacts can be made only to electrons confined near to surface where mobility is dominated by surface scattering. Depending on the application, in some cases high electron mobility is desired \cite{Sau12,Haim2018} and in other cases control over the induced gap is called for \cite{Cole15}. It was determined in near surface 2DEGs, that a 10~nm thick top layer of In$_{0.81}$Ga$_{0.19}$As can achieve both \cite{Shabani2016,Kaushini2018}.%In case of Al-InAs structures, signatures of Majorana fermions have been observed \cite{Henri17,Fabrizio17}.

%on 2D hybrid InAs-Al system and HgTe proximitized by Al and Nb \cite{Hart2014, Molenkamp}.

%Recently the interest in material synthesis of InAs structures has been renewed mainly due to developments in engineering novel quasiparticles, Majorana zero modes, in semiconductors with strong SOC (InAs, InSb) in the proximity from conventional superconductors (Al, Nb)  \cite{Mourik2012,Deng16,Gul18}.  In order to establish their statistics and eventually move beyond demonstrations of zero bias signatures to braiding \cite{AliceaNatPhy11, HalperinPRB12} and larger-scale Majorana networks \cite{AliceaBraiding,Matos} it is likely that a top-down patterning approach will be needed.  Growth of large-area 2D Superconductor-Semiconductor (S-Sm) systems through Molecular Beam Epitaxy (MBE) can provide the basis for such an approach. 

In this work, we study the electronic transport properties and their connection to Josephson effect properties in InAs structures with Al epitaxial contact. We study these properties in samples of different bare mobilities (not gated) to not only compare the nature of the interface formed by in-situ epitaxial growth but also the effect of the 2DEG quality. These hybrid systems can be characterized by study of the $I_cR_N$ and $I_{ex}R_N$ products, where $R_N$ is the normal resistance of the JJ. $I_{ex}$ is the difference between the measured current through the junction and the expected current based on the junctions $R_N$. This occurs due to Andreev reflections and depends primarily on interface transparency. $I_{c}$ is the amount of current that can be carried by Andreev bound states through the junction with zero resistance. $I_{c}$ requires coherent charge transport across the semiconductor region, so it is a measure of both interface transparency and 2DEG mobility. Investigating both products provides a means of studying the effects of both interface transparency and 2DEG mobility. We find that 2DEG mobility and the inferred interface transparency seem to be closely related, possibly due to the fact that the mobility of surface 2DEGs has been found to be dominated by surface scattering  \cite{Kaushini2018}. This implies that the same impurities affecting the 2DEG mobility will also dominate degradation of the interface in the case of in-situ epitaxial growth, which would otherwise produce a transparent interface. In higher mobility samples we study $I_cR_N$ for different junctions lengths, temperatures, and gate voltage where the analysis is in agreement with a fully transparent metallic interface. We also find that $I_{ex}R_N$ is independent of JJ length and applied gate voltage in contrast to previous studies \cite{Abay2014,Li2016,Samuelsson2004}.  In our high mobility 100~nm short junction we report $I_{c}R_{N}/\Delta$ values up to 2.2 and $I_{ex}R_N/\Delta \sim 1.5$.

%\section{sample Growth and Fabrication}

The samples were grown on semi-insulating InP (100) substrates. The quantum well consists of a 4 nm layer of InAs grown on a 6 nm layer of In$_{0.81}$Ga$_{0.19}$As. Since coupling the 2DEG to a superconductor is the main requirement, the charge distribution at the semiconductor metal interface has to be finite. To satisfy this condition we grow a 10nm In$_{0.81}$Ga$_{0.19}$As layer on the InAs which has been found to produce an optimal interface while maintaining relatively high 2DEG mobility \cite{Kaushini2018}. After the quantum well is grown, the substrate is cooled to promote the growth of epitaxial Al (111). Wafer A and B are grown with identical nominal structure. The difference in arsenic overpressure (or equivalently III/V ratio) during the growth of lattice-mismatched buffer results in different misfit dislocations and varied roughness.  Atomic force microscopy images of samples are shown in Fig. 1(a) and 1(b). Images show variation of minimum and maximum topography within a 34 $\mu$m square window to be approximately 9~nm for sample A and 7~nm for sample B. It should be noted that these images are taken from the full structure with top Al layer. The roughness is similar when Al is selectively removed indicating roughness is due primarily to underlying semiconductor structure, not Al.

To determine the mobility and density, selective wet etching techniques are used to remove the top Al layer to measure both longitudinal and transverse resistances in a van der Pauw geometry. The magnetoresistance for the two samples we will consider in this paper are plotted in Fig. 1(c) and 1(d). Josephson junctions are fabricated with electron beam lithography. Devices are gated using 50~nm of AlO$_x$ deposited by ALD followed by lithographically defined Ti/Au gates.  All studied junctions have a 4~$\mu$m width (W) with varying gap lengths (L) as shown in Fig.~2(a) and are measured in a 4-point geometry. Measurements are performed in a dilution fridge with mixing chamber temperature of 7~mK and an estimated electron temperature of 20~mK.

%\section{Interface and 2DEG Characterization }

When considering the properties of these InAs JJ's, both transparency of the InAs-Al interface and scattering within the 2DEG need to be considered.  Scattering in the 2DEG determines whether transport through the junction is diffusive or ballistic, generally characterized by the mean free path  $l_{e} $ in comparison with junction length, $L$. Magnetotransport for samples shown in Fig. 1 yields mean free paths $l_{e}^A= 87$~nm and $l_{e}^B= 201$~nm, where superscript denotes the sample.. We fabricate multiple junctions on samples taken from wafer A and B. With JJ lengths ranging from 50~nm to 1~$\mu$m, mean free paths of this order indicate transport is between the diffusive and ballistic limits. The $T_c$ of both Al thin films is measured and found to be $T_c^A=1.53$~K and $T_c^B=1.48$~K Using the relation $\Delta_{Al}= 1.75k_BT_c$ we find $\Delta_{Al}^A=231$~$\mu$V and $\Delta_{Al}^B=223$~$\mu$V with the variation being attributed to slightly different Al film thicknesses. This is supported by perpendicular magnetic field measurements yielding critical fields of $B_c^A=164$~mT and $B_c^B=96$~mT. From $\Delta_{Al}$ we can estimate the superconducting coherence length in our samples given by $\xi_0=\hbar v_F/(\pi \Delta)$, which yields  $\xi_0^{A} =635$~nm and $\xi_0^{B}=774$~nm for respective samples. From this we expect all devices to approach the dirty limit($\xi_0\gg l_e$). This implies we should also consider the dirty coherence length  $\xi_{0,d}=\sqrt{\xi_0 l_e}$ which yields $\xi_{0,d}^{A}=235$~nm and $\xi_{0,d}^{B}=394$~nm. A summary of devices fabricated on both wafer A and B are shown in Fig.~2(b). 

\begin{figure}[t!]
\centering
\includegraphics[scale=0.5]{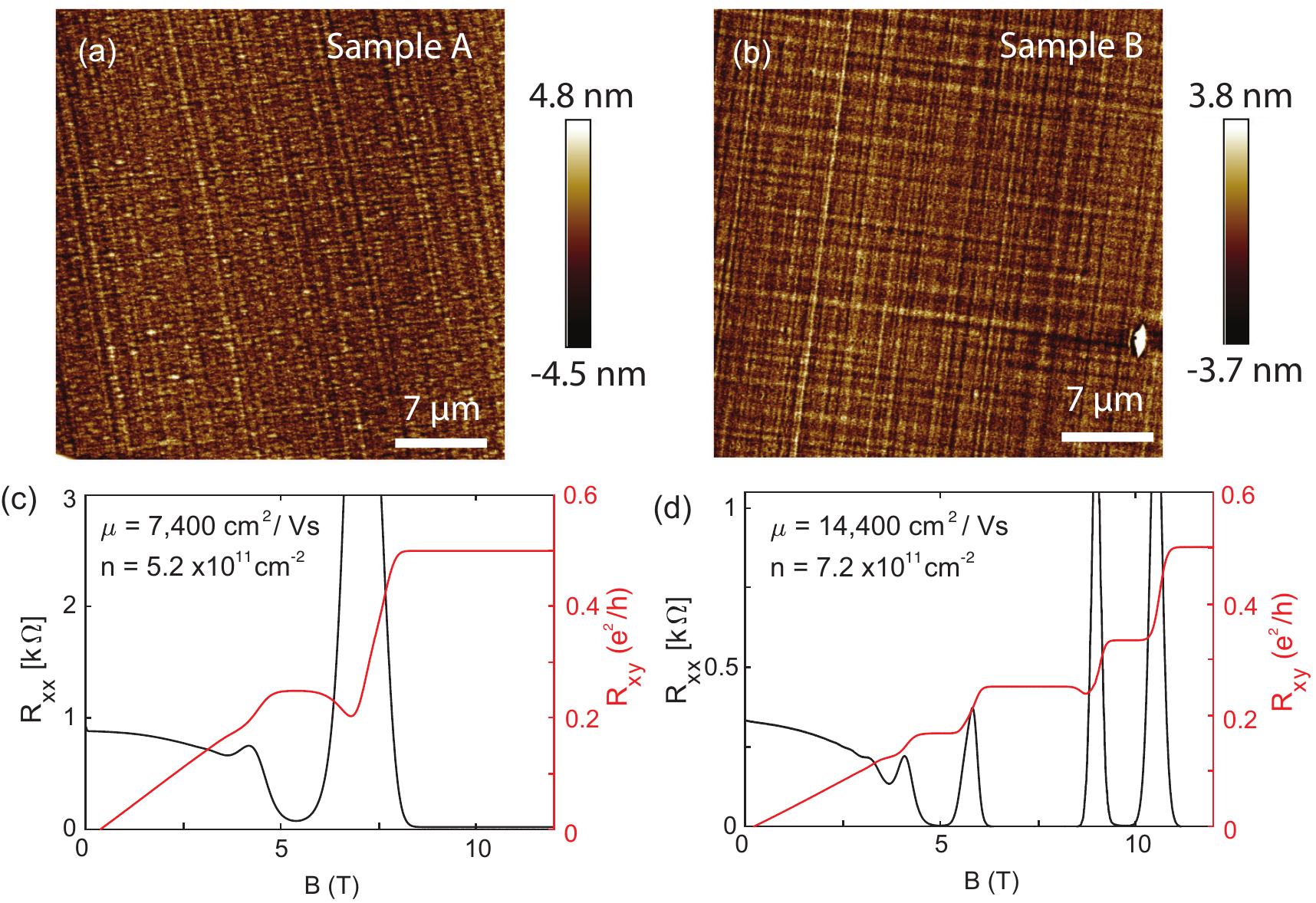}
\caption{(Color online) Atomic force microscopy image of (a) sample A and  (b) sample B.  (c) Longitudinal (black) and Hall (red) resistance shown as a function of magnetic field for sample A and (d) similarly for sample B at 20~mK.}% (e,f) VI curve for 100nm Al-InAs JJ on sample A and B at 20mK. Dashed line shows linear extrapolation of slope at $V\gg\Delta_{Al}$ to zero voltage yielding $I_{ex}$.}
\end{figure}

The standard figure of merit for JJ is $I_cR_N$. It should be noted that experimentally there is a distinction between $I_c$ and the measured switching current of a junction. In an ideal system these quantities would be identical but it is commonly seen that finite temperature or noise can cause the junction to prematurely switch. This leads to measuring an observed switching current lower than $I_c$. For simplicity in the analysis, we assume that they are equal in our devices while knowing this can lead to an underestimation of our $I_c$.  The critical current can be related to the gap by the formula $I_{c}R_{N} = \alpha \Delta_{0}/e$. From Fig.~3 (a,b) insets we find $I_{c}R_{N}^A = 135$~$\mu V$  and $I_{c}R_{n}^B= 486$~$\mu V$. These values can be compared to theoretical values for fully transparent junctions in the short diffusive and ballistic limits, for which $\alpha$ are $1.32(\pi/2)$ and $\pi$ respectively \cite{IcRnpaper,IcRnreview} when JJs are deep in ballistic ($l_e \gg L$) or diffusive regime ($L \gg l_e$). For sample B we find $I_{c}R_{N} $ is $69\%$ of the ballistic limit and $105\%$ of the diffusive limit. In contrast the results for sample A are $17\%$ of the ballistic limit and $28\%$ of the diffusive limit. From Fig.~2(b) it is clear that 100~nm JJ in sample B is closer to the short ballistic regime while 100~nm JJ in sample A is closer to the short diffusive regime. This is supported by $I_{c}R_{n} $products where for sample B the product exceeds the diffusive limit indicating its ballistic character. In contrast sample A $I_{c}R_{n} $ product is well below even the diffusive limit indicating diffusive transport is most likely. 

High interface transparency corresponds to a high probability of Andreev reflection at the interface. Since the Sm extends under the S regions, the interface between Sm and S should be highly transparent due to the large area of contact and in-situ epitaxial Al growth \cite{MortenNatureComm16}. The Andreev process that carries the supercurrent across the Sm region is characterized by the excess current($I_{ex}$) through the junction $I_{ex}= I-V/R_N$ \cite{Tinkham1982}. Excess current does not require coherent charge transport across the junction as it follows simply from charge conservation at the S-Sm interfaces. This allows for the excess current to be calculated by extrapolating from the high current normal regime to zero voltage as shown in Figures 3(a) and 3(b) with dotted lines. The excess current in samples A and B are found to be $I_{ex}^A=20$~nA and $I_{ex}^B=3.5$~$ \mu  A$ respectively for 100~nm JJ.

When considering interface quality the more relevant quantity is the product $I_{ex}R_N$. The product $I_{ex}R_N$ can be compared to the superconducting gap with the relation $I_{ex}R_N=\alpha'\Delta_0/e$. In the case of a fully transparent S-Sm interface $\alpha'=1.467$ for a diffusive junction \cite{IexeRnpaper} and $\alpha'=8/3$ for a ballistic junction \cite{Tinkham1982}. For samples A and B, $I_{ex}R_N^B=30$~$\mu V$ and $I_{ex}R_N^A=340$~$\mu V$. Comparing these to the ballistic and diffusive limits we see that our values are $57\%$ of ballistic limit and at $104\%$ actually slightly exceeds the diffusive value for a 100~nm JJ on sample B. Sample A is at only $5\%$ for ballistic and $9\%$ of diffusive limit. As indicated previously, for $L=100$~nm the ratio $L/l_e$ is close to unity. Consequently for a fully transparent interface we would expect $I_{ex}R_N$ to be between the two limiting cases as is observed for sample B. 

\begin{figure}[t]
\centering
\includegraphics[scale=0.45]{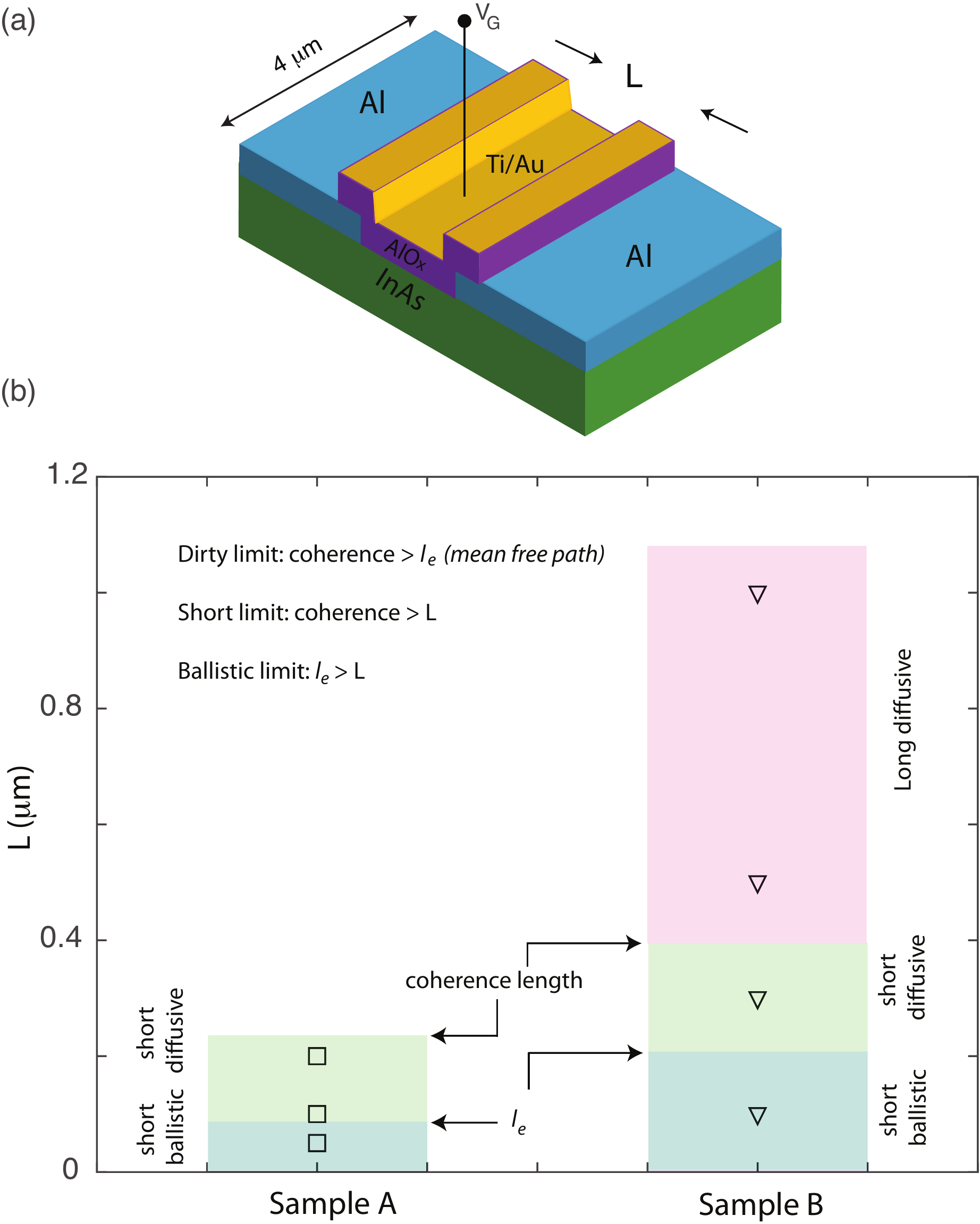}
\caption{(Color online) (a) Schematic of gated JJ. (b) Summary of devices fabricated, indicated by squares (sample A) and triangles (sample B), from two wafers . The relevant length scale are junction length, L, coherence length and mean free path, $l_{e}$. All samples are in the dirty limit.}
\end{figure}

While the mobilities of the samples only differ by a factor of two, both figures of merits, $I_{c}R_N$ and $I_{ex}R_N$, indicate a starker contrast between the two samples. Additionally the difference is more pronounced in the case of $I_{ex}R_N$ product which depends primarily on interface transparency. This leads us to conclude that the interface transparency is different in the two samples. We should note that in near surface 2DEGs it could be that interface transparency and mobility are coupled. It has been shown in these materials that the dominant scattering mechanism is surface scattering \cite{Kaushini2018}. If scatterers are present on the surface they will also affect the interface to the metal, making mobility a good proxy for interface transparency, at least for the ranges of mobility studied in this paper.

%$l_{e}$ is related to the superconducting gap by the relationship $l_{e} R_N= \alpha\Delta/e$ $\Delta_{\rm ind}$ in the Sm below the S rather than the bulk Al gap, $\Delta_0$.  

The values for $I_{ex}R_N$ and $I_{c}R_N$ indicate sample B has both a highly transparent S-Sm interface and a high quality 2DEG which can support coherent transport in a 100~nm JJ. Alternatively sample A has drastically lower values despite the same in-situ epitaxial contacts. So while both samples show robust DC Josephson effect made possible by in-situ epitaxial contacts, this result leads us to conclude that in-situ epitaxy does not necessarily lead to transmission near unity.
This emphasizes the importance of growing high mobility surface quantum wells for applications that require transparent S-Sm interfaces such as the search for topological superconductivity. 

% and depends on scattering present on the surface. 

\begin{figure}[t]
\centering
\includegraphics[scale=0.5]{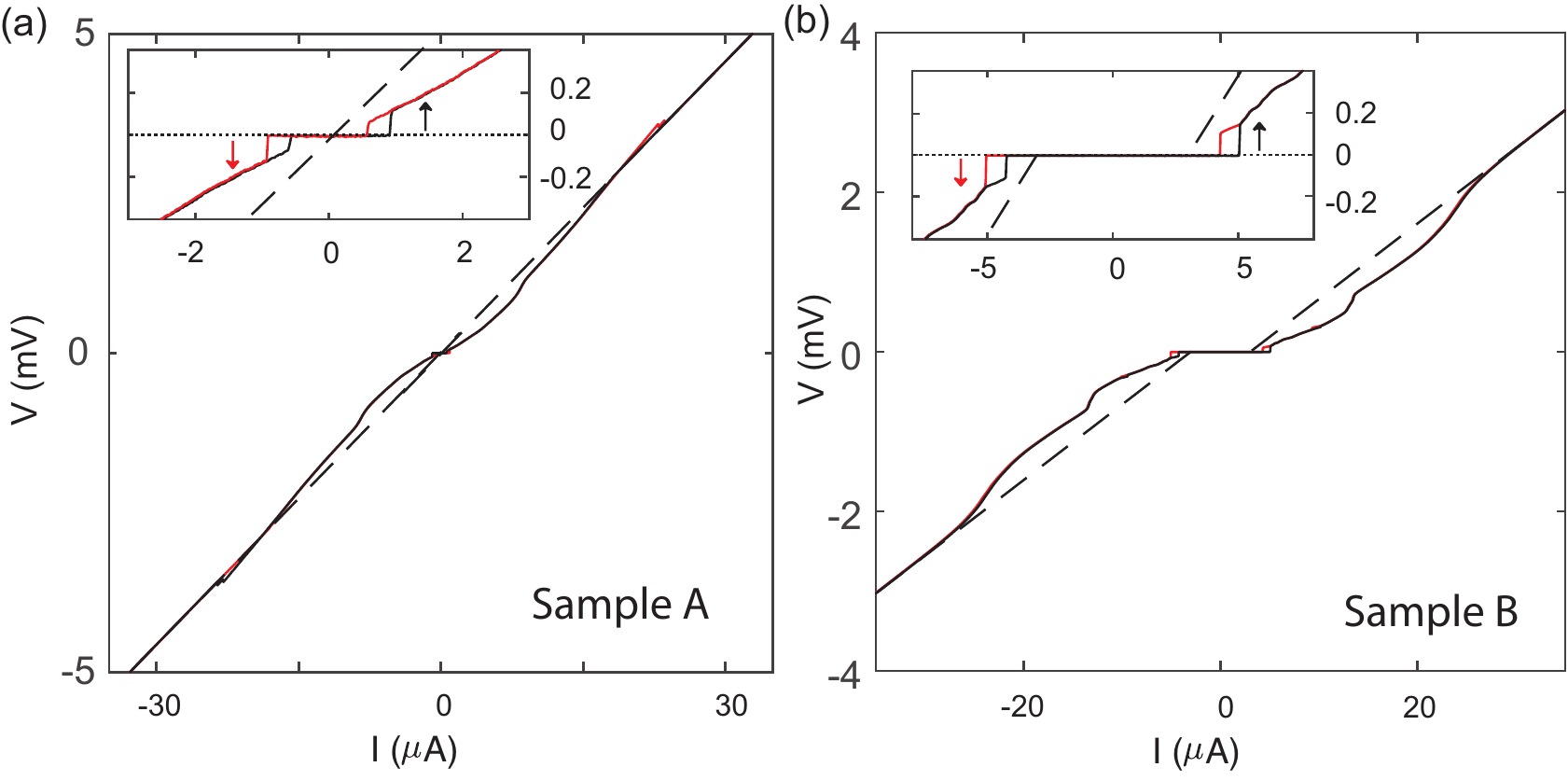}
\caption{(Color online)  Voltage-current curve for 100~nm Al-InAs JJ (a) on sample A (b) and sample B at 20~mK. The inset shows near zero bias data where the current switches. Higher bias data show the  linear extrapolation from normal state (crossed checked with finite magnetic field) to zero voltage yielding $I_{ex}$.}
\end{figure}

%The fact that the mobility in the two samples is different by approximately a factor of two and density is slightly lower in sample A can't account for such a dramatic difference. So while both samples show robust dc Josephson effect made possible by $in-situ$ epitaxial contacts, this result leads us to conclude that even with $in-situ$ epitaxy the interface is not necessary transparent due to scattering present on the surface of the lower quality 2DEG. The nature of this surface scattering is not well understood and generally simply modeled as remote impurity doping \cite{Kaushini2018,Gold87}. This emphasizes the importance of growing high mobility surface quantum wells for applications that require transparent S-Sm interfaces. 

%Previous studies of ex-situ fabricated junctions on 2DES have typically reported $I_cR_n$ products an order of magnitude smaller than $\Delta_{0}$ \cite{Nitta92,Takayanagi95,Mur96,Heida98}. 

%\section{Josephson Properties as Function of Junction Length}

The length of the 2DEG channel between superconducting electrodes, $L$, can be varied as shown in Fig.~2(b). Figure 4(a) shows the products $I_{c}R_N$ for both samples on junctions up to 1~$\mu m$ using cross symbols. $I{_c}R_N$ decreases with junction length. From the theory for long junctions where $I_cR_N=I_{c0}R_Ne^{-(L/\xi_{0,d})}$ \cite{Delin1996}, we plot the theoretical expectation as the solid line (in range of 400~nm up to 1~$\mu m$) with prefactor (zero length intercept) found to be $I_{c0}R_N \sim 900$~$\mu$V while for small values of $L$ we expect $I_{c}R_N=\pi\Delta_{0}/e \sim 700$~$\mu$V, the short ballistic limit. 

\begin{figure}[htp]
\centering
\includegraphics[scale=0.35]{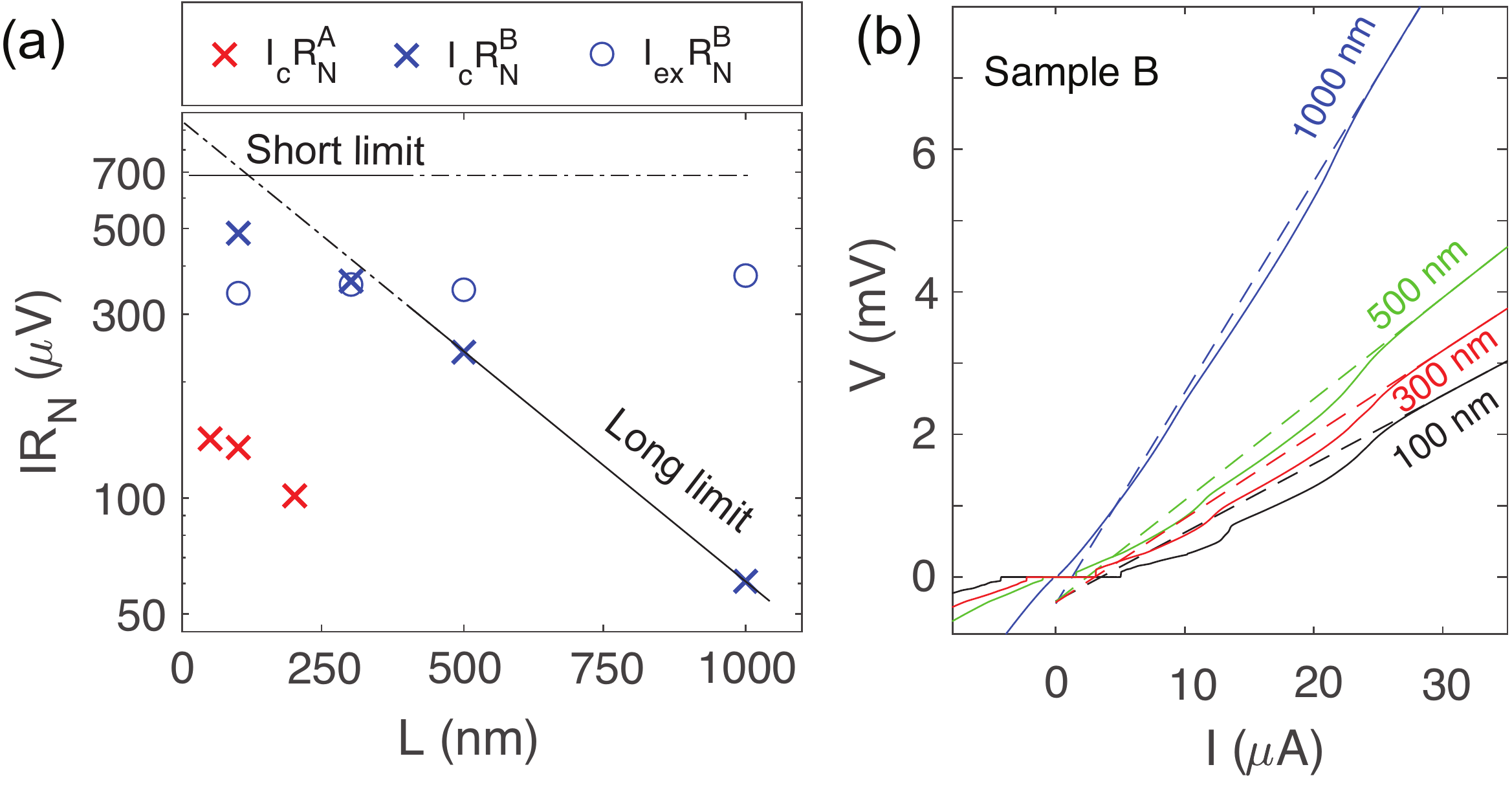}
\caption{(Color online) (a)  Semi-log plot of dependence of $I_{c}R_{n}$ on junction length for sample A (red crosses) and sample B (blue crosses) compared to theoretical exponential dependence (solid black line). Also shown are values of $I_{ex}R_{n}$ (filled blue circles) as a function of junction length for sample B.  (b) VI curves for JJ on sample B of various lengths. Dashed line shows linear extrapolation of slope at high biases to zero voltage yielding $I_{ex}$. }
\end{figure}
%Normal resistivity per nm for $W = 4 \mu$m wide junctions as a function of length. Solid lines indicates estimated Sharvin resistance for sample B. (c) 

 Figure 4 also shows $I_{ex}R_N$ for sample B. The values for sample A(not shown) are trivially constant. As previously stated for this sample even at $L=100$~$nm$, $I_{ex}=20$~$nA$ and remains small for all measured lengths. In contrast with $I_{c}R_{n}$ length dependence we see no significant change in $I_{ex}R_N$ over the studied lengths \cite{Abay2014,Li2016,Samuelsson2004}. This can also be seen in Fig. 4(b), since the y-intercepts of the extrapolated linear fits to the VI curves are simply $I_{ex}R_N$.  A decrease in $I_{ex}R_N$ with JJ length has always previously been observed experimentally in both diffusive and ballistic junctions on various materials. It has been shown theoretically that in ballistic junctions $I_{ex}R_N$ will only remain constant with junction length for highly transparent interfaces \cite{Ingerman2001}. %While there is no formal theory to the authors knowledge, it seems reasonable to extended this expectation to the diffusive limit. This would match with experimental results in sample B since junction range from quasi-ballistic($L \sim l_e$) be well into the diffusive regime.

%The JJ normal state resistance per unit length for both samples is shown in Figure 2b. 
%For longer junctions resistance per unit length in sample B appears to become a constant as expected in the diffusive regime. Both samples diverge from this dependence at shorter lengths. In the ballistic limit $R_N$ should be described by a number of modes each carrying $h/(2e^2)$ resistance. The number of channels is related to the ratio of the junction width(W) and the Fermi wavelength($\lambda_F$). So the resistance of such a junction can be roughly estimated by the Sharvin resistance, $R_{sh}=h/(2e^2)\lambda_F/2W=48\Omega$, here calculated for sample B. The ratio $R_{sh}/L$ is plotted in Fig.~2(b) in black line. Comparison with this curve indicates for shorter junctions the resistance is best described by a mix of ballistic and diffusive resistances, consistent with the values of $I_cR_N$ and $I_{ex}R_N$ measured in sample B.

%\section{Temperature Dependence of $I_cR_N$}

The product $I_cR_N$ can be directly varied by changing the Al superconducting gap by increasing temperature. Figure 5a shows the temperature dependence of $I_cR_N$ of 100~nm JJs for sample A and B. The critical temperature of  in-situ Al thin films are slightly enhanced over the bulk value of 1.2 K to near 1.5 K, due to thickness \cite{Tedrow82, Shabani2016}. The temperature dependence can be fitted with the generalized Kulik-Omelyanchuk relation \cite{Delin1996} where

\begin{eqnarray}
I_cR_N= \frac{\alpha\Delta(T)}{2e}\frac{sin(\phi)}{\sqrt{1-\tau sin^2(\phi/2)}} \\
\times tanh\Big[\frac{\Delta(T)}{2k_BT}\sqrt{1-\tau sin^2(\phi/2)}\Big] \nonumber
\end{eqnarray}

The actual $I_cR_N$ is found by maximizing $I_cR_N(\phi)$ with respect to phase $\phi$. $\Delta(T)$ is the BCS gap calculated using $T_c$ of Al and $\tau$ is a measure of interface transparency with $\tau=1$ being fully transparent. In the limit $T=0$ and $\tau=1$ we recover the relationship $I_cR_N= \alpha\Delta/e$. The equation fits sample B data best for $\alpha=0.69\pi$ and $\tau=1$ consistent with the values found from just the 100 nm JJ at 20mK. Thus temperature dependence of sample B also indicates a highly transparent interface.

The JJs are equipped with gates that allow the junction resistance to be varied up to a fully insulating state. Figure 5(b) shows the dependence of products $I_cR_N$ and $I_{ex}R_N$ on gate voltage for the 100~nm JJ on sample B. The inset shows the dependence of $R_N$ on gate voltage found from slope of IV curves at high current (black crosses) and differential resistance measured at $B_{\perp}=110$~$mT$ where Al is no longer superconducting. In the regime of high resistance and large current, nonlinearities can affect the extraction of $R_N$ from the slope of IV outside the Al gap. The reliability of the extrapolation in the range plotted is confirmed by comparison of the extracted normal state resistance with resistance from gate voltage sweeps at $B_{\perp}=110$~$mT$. While the density at zero gate voltage is taken from magnetotransport measurements on the same sample the JJ geometry does not allow for a dependable measurement of density for non-zero gate voltages. Previous studies of InAs 2DEG based devices have shown a small increase in $I_cR_N$ with gate voltage \cite{Shabani2016,MortenPRApplied17}. This feature is not present in our junction possibly due to the lower initial density of our samples. We observe a decrease of $I_cR_N$  with applied gate voltage with $I_c=0\mu A$ occurring at $R_N=2.7$~$k\Omega$.

%\begin{figure}[htp]
%\centering
%\includegraphics[scale=0.63]{fig4_newAPL-eps-converted-to.pdf}
%\caption{ Gate dependence of products $I_cR_N$ and $I_{ex}R_N$ are shown for 100nm JJ on sample B. Inset shows gate dependence of junction $R_N$ for same device. Black crosses show $R_N$ from high current VI slope. Red line shows $R_N$ measured at $B_{\perp}=110mT$ where Al is not superconducting.}\end{figure}

\begin{figure}[t]
\centering
\includegraphics[scale=0.625]{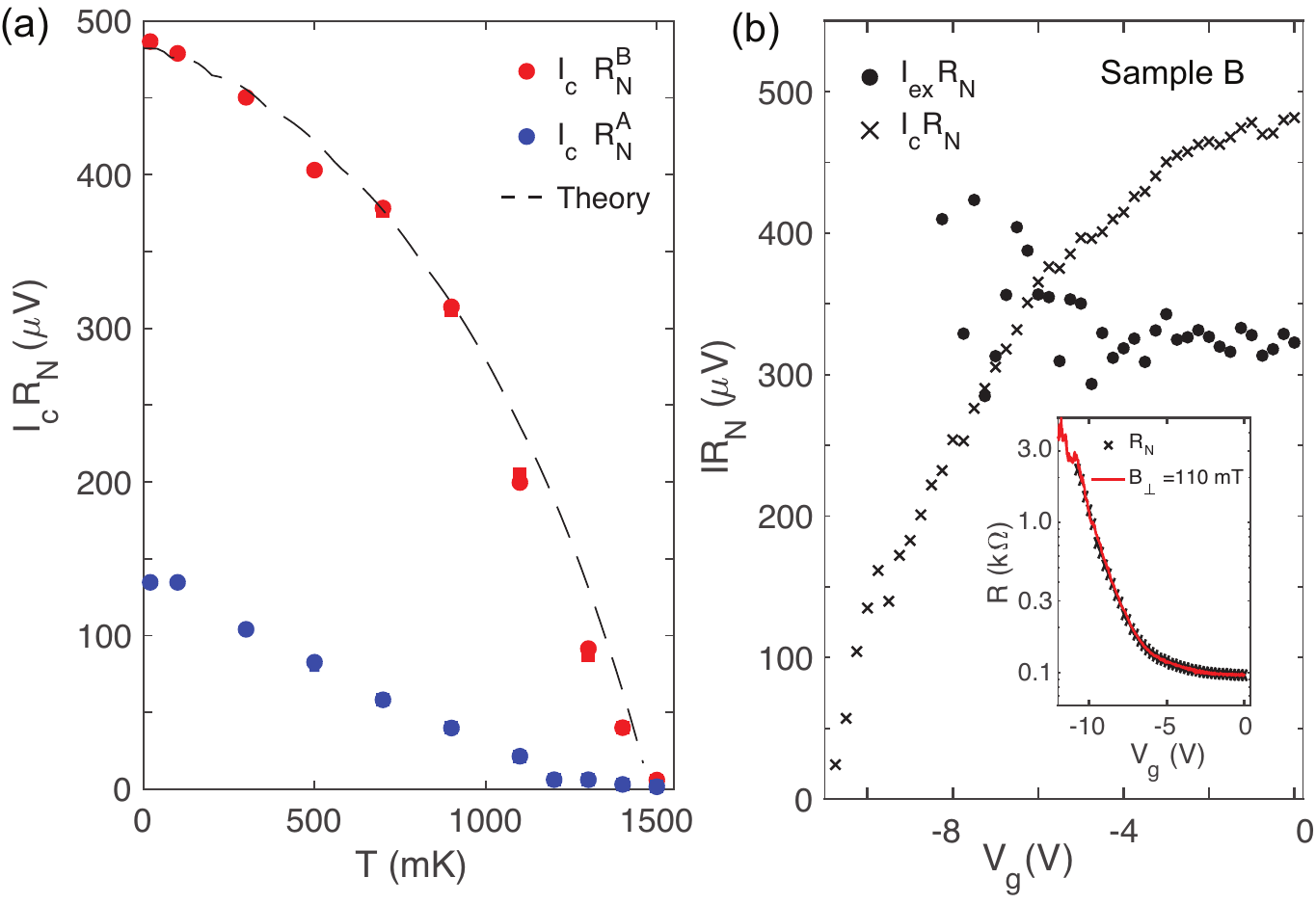}
\caption{ (Color online) (a) Temperature dependence of the $I_cR_N$ products for both samples. Dashed line indicates fitting $I_{c}R_N^B$ data with unity transparency formula, see text. (b) Gate dependence of products $I_cR_N$ and $I_{ex}R_N$ are shown for 100~nm JJ on sample B. Inset shows gate dependence of junction $R_N$ for same device. Black crosses show $R_N$ from high current VI slope. Red line shows $R_N$ measured at $B_{\perp}=110$~$mT$ where Al is not superconducting.}
\end{figure}

	 In contrast to the decrease in $I_cR_N$, $I_{ex}R_N$ does not change appreciably with gate voltage, as previously observed with junctions of different lengths on sample B.  At large negative gate voltages $I_{ex}$ becomes very sensitive to noise making extrapolation of the product $I_{ex}R_N$ difficult for high resistances. Consequently the most negative gate voltage used for measurement is $V_g=-8V$ where $R_N=300$~$\Omega$. The product $I_{ex}R_N$ at $V_g=-8$~$V$ is found to be very similar to $V_g=0$~$V$ despite having three times the normal resistance. At $V_g=-8$~$V$, $I_cR_N$ is about half it's value at $V_g=0$~$V$.  This gate voltage independence further emphasizes that $I_{ex}R_N$ depends primarily on interface transparency which is unaffected by density changes in the 2DEG.

In conclusion we study Josephson junctions with in-situ epitaxial contact to InAs 2DEG's with different mobilities. We observe a remarkable difference in junction properties $I_cR_N$ and $I_{ex}R_N$. Since both samples have in-situ epitaxial contact this difference is unexpected. One possible explanation is surface scattering, which is the primary scattering mechanism for these structures, contributing an extra component to scattering at the interface. These results indicate that just having in-situ epitaxial contact does not guarantee a transparent interface. %The exact details of interface scatterers are unknown at this point but seems to correlate with electron mobility.
In the higher mobility sample for 100~nm JJ we find product $I_cR_N/\Delta_{0} \sim2.2$. Remarkably we observe $I_{ex}R_N \sim 1.5$ to be independent of both junction length and gate voltage in this sample. This is in contrast to previous studies which all see a decrease in $I_{ex}R_N$ and is the strongest indication of the highly transparent interface from in-situ epitaxial contact to a high mobility surface 2DEG.

%The change in $I_cR_N$ with junction length in long junction is consistent with the expected exponential decay with a highly transparent interface. Temperature dependence is also in agreement with predicted dependence for a transparent interface. 

This work was partially supported by US Army research office, Intel and DARPA Topological Excitations in Electronics (TEE) program. We thank Vladimir Manucharyan for illuminating discussions.

\bibliography{References_Shabani_Growth}

\end{document}